# Temporal sequence of deformation twinning in CoCrNi under tribological load

Antje Dollmann[1,2], Julia Rau[1,2], Beatrix Bieber[1], Lakshmi Mantha[3,5], Christian Kübel[3,4,5], Alexander Kauffmann[1], Aditya Srinivasan Tirunilai[1], Martin Heilmaier[1], Christian Greiner[1,2]

[1] Institute for Applied Materials (IAM), Karlsruhe Institute of Technology (KIT), Kaiserstraße 12, 76131 Karlsruhe, Germany

[2] IAM-ZM MicroTribology Center (µTC), Straße am Forum 5, 76131 Karlsruhe, Germany

[3] Institute of Nanotechnology (INT), Karlsruhe Institute of Technology (KIT), 76344 Eggenstein-Leopldshafen, Germany

[4] KIT-TUD-Joint Research Laboratory Nanomaterials, Technical University Darmstadt (TUD), 64287 Darmstadt, Germany

[5] Karlsruhe Nano Micro Facility (KNMFi), Karlsruhe Institute of Technology (KIT), 76344 Eggenstein-Leopoldshafen, Germany

Abstract

Microstructural evolution under tribological load is known to change the friction and wear response of the entire tribological system. It is however not yet fully understood, how these changes take place on an elementary, mechanistic level. Revealing the temporal sequence of deformation mechanisms under a tribological load by experiments is scarce. The discrete strain release in the case of deformation twinning allows the identification of their temporal sequence. Therefore, a medium stacking fault energy material, namely CoCrNi, was investigated due to the significant contribution of deformation twinning to the overall deformation at room temperature. The investigated grain showed three activated twin systems. The evolution sequence of these is discussed critically based on several twin intersection mechanisms. The experimental analysis decodes the temporal sequence of the active twinning systems of highest resolved shear stresses for the first time. This approach gives experimental indications for the stress field under a tribological load.

Graphical Abstract

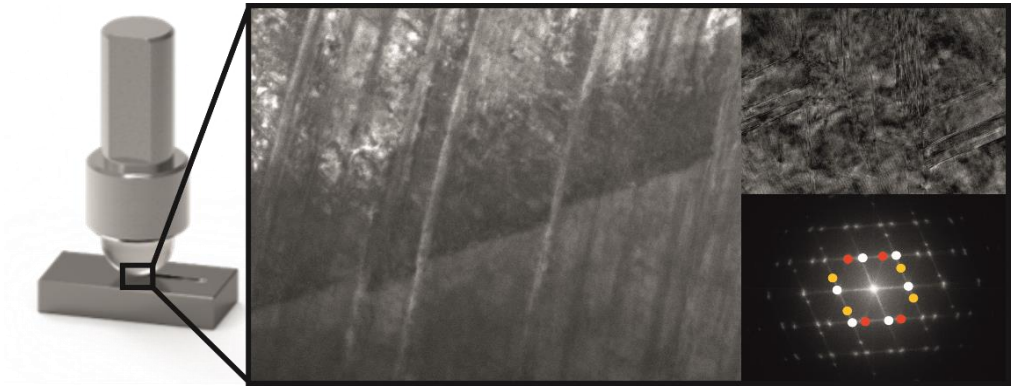

Main body

Tribological loading of metals and alloys leads to microstructural changes in the subsurface region [1,2] that alter the mechanical response of the material. Consequently, it is not surprising that this microstructural evolution has an influence on the friction and wear behavior [3]. So far, predicting the microstructural evolution is challenging although a wide range of experimental parameters, e.g. material pairing [4–6], sliding velocity [7,8], normal force [4,9], atmosphere [10] and temperature [8,10], have been investigated. The microstructural evolution has been characterized *ex situ* subsequent to the tribological experiment. As a result, subgrain formation [5,6], twinning [8,11,12] and nanocrystalline grains [5,7] have been found. Attempts have been made to understand the microstructural evolution by investigating the microstructure after different number of sliding cycles [5,6,11,13]. This approach gave insights into how the involved subsurface region grows and how the (sub-)grain size changes [13]. It was possible to identify different strain components in the vicinity of an annealing twin boundary, namely simple shear, localized shear and the accommodation of crystal rotation by forming a grain boundary parallel to the surface (called dislocation trace line (DTL)) [14]. A model of the dislocation motion based on the Hamilton stress field was used to explain the formation of DTLs [15]. However, knowledge about the activated slip and twinning systems and the sequence of their activation is scarce in literature.

From an experimental point of view, *in situ* tribological tests probing the microstructural evolution are challenging and were for the authors' best knowledge only carried out on olivine (magnesium iron silicate) [16]. Molecular dynamics (MD) simulations have been conducted [17,18], which can accurately track the temporal sequence. However, the modelled volumes are small, the applied sliding velocities are high in comparison to experiments and the sampled grain sizes are much smaller than those of engineering materials.

In this study, we performed an experimental analysis of the temporal sequence of deformation twinning under a tribological load. Twinning as an elementary mechanism has the advantage for such a study that due to its discrete strain release on a limited set of crystallographic planes in specific directions and strict unidirectionality, it can be more clearly analyzed than dislocation motion. Deflection of twin laths is impossible in contrast to cross-slip of dislocations making the post-mortem analysis of twins less challenging. For twin formation, the critical shear stress for twinning has to be reached which is higher than for dislocation glide at room temperature [19]. The resolved shear stress is dependent on the time-varying stress field under tribological load and latent hardening.

We rely on twin intersections to determine the temporal sequence. For this reason, the activation of at least two twinning systems is needed, which requires a material and grain orientation favoring deformation twinning. In order to achieve such circumstances, polycrystalline coarse-grained CoCrNi was used. CoCrNi has a stacking fault energy of $(22 \pm 4)$ mJ/m$^2$ [19] and tensile tests on polycrystalline samples have shown twin formation on several systems at room temperature [20,21].

The details of the material synthesis, the sample preparation and high-resolution transmission electron microscopy (HR-TEM) analyses are given in the supplementary material. The subsurface microstructure before the experiment is presented in Figure S1. A self-built tribometer was used for the tribological experiments [13]. It was encapsulated in a climate chamber to ensure a dry $N_2$ atmosphere for a single stroke experiment without lubrication. The experimental parameters were a normal load of 2 N, sliding velocity of 0.5 mm/s, sliding distance of 12 mm and a SiC sphere counter body with 10 mm diameter was used (hightech ceram, Dr. Steinmann + Partner GmbH, Germany). The friction coefficient over sliding distance, electron microscopy images of the wear track and the counter body are given in Figure S2. After the single stroke experiment, material transfer is observed on the SiC sphere as well as flake formation in the wear track, both indicative of adhesive wear. The friction coefficient is ~0.3. The sliding distance in Figure S2a at 6 mm is marked, as at this position the microstructure was investigated by preparing a TEM foil.

In Figure 1a, an overview scanning transmission electron microscope (STEM) image of the entire thinned region is presented. Three band line-like features are observed. The transmission Kikuchi diffraction (TKD) measurement in Figure S3a demonstrates that all of these features correspond to twins. This means that the requirement of at least two activated twin systems for analyzing the temporal sequence is fulfilled. The twin system TS1 colored in orange has an inclination angle of 25° to the surface, tilted in sliding direction (SD). The most frequently observed twin system TS2 has an inclination angle of around 70° to the surface, slightly bending along SD close to the surface. This system is color-coded in red. The twin system TS3, colored in green, occurs beneath the orange-colored twins, has an inclination angle of 27° to the surface and is inclined with respect to SD. The twin plane normal analyses are given in Figure S3b-d.

The names and colors for the different twin systems are used throughout the manuscript. Higher magnification micrographs of twin intersections of TS2 and TS3 as well as TS1 and TS2 are given in Figure 1b+c. The twin intersection in Figure 1b shows a straight twin on TS3 and a displaced twin on TS2. Figure 1c displays a straight twin on TS2 intersecting with a displaced twin on TS1. It should be noted that the offset of the twin on TS1 in Figure 1c is in different direction in comparison to TS2 in Figure 1b.

Revealing the temporal evolution needs the identification of the barrier twin (BT; twin, which was formed first) and the incident twin (IT; twin, which was formed second) in the twin intersections. In literature, twin intersections differ by their appearance (Figure S4). Intersections where both twins are displaced [22–24] exist as well as intersections, where one twin is straight and the other is displaced [25,26]. In Figure 1 to Figure 4, all intersections exhibit one straight twin. For this reason, this type will be solely considered in the following. There is no consensus in the literature concerning the temporal sequence of the two twinning systems [22,23,26,27]. Several approaches have been taken to determine the temporal sequence: (1) the IT has to be continuous in a dark-field TEM image [27,28], (2) twins on one system can either be intersecting or be stopped by a twin on another system; in this case, the stopped twin system is the IT [22,29]. For both cases, contradicting formation mechanisms exist. The authors of Refs. [24] and [30] propose that the IT crosses the BT by perfect dislocations, which does not change the crystal orientation of the BT in the intersection region. Furthermore, Zhang and Ye [25] suggest that the shear is on a (511) 1/18[$\bar{1}\bar{7}2$] system in the BT. Such a shear results in an AA stacking sequence. An atomic shuffle is supposed to reorient the lattice into the orientation of the BT. Both attempts result in a continuous contrast for the BT under dark-field conditions. A contradicting mechanism for (2) was given by Müllner and Solenthaler [31]. Here, the BT is partly detwinned, which has afterwards the same appearance as a twin stopping at another one.

In any case, a BT has to be displaced after being crossed by an IT as all twins exhibit a shear operation. This assumption as well as approach (1) and (2) will be used in the following. In Figure 2a, a HR-TEM image of the intersection between the TS1 and TS2 is displayed. Figure 2b shows the same image, but with additional labels and Figure 2c displays the corresponding fast Fourier transformation (FFT). The filtered inverse FFT (iFFT) images of TS1 and the TS2 are given in Figure 2d+e. In the iFFT images, it can be clearly seen that the twin on TS2 is straight and that the twin on TS1 is displaced. This means that the twin on TS2 is the IT while the twin on TS1 is the BT. A possible twin intersection mechanism for this scenario was given by Müllner [26]: Full dislocations glide parallel to the BT boundary. If this mechanism is considered on an atomic level (Figure S5a), it is not obvious how the glide of perfect dislocations rotates the crystal of the BT in the crystal orientation of the IT. The boundary limiting the intersection zone in the model is not observed in Figure 2. In general, uncertainties exist with most of the published intersection mechanisms, as it is not known which stresses activate the proposed reactions [22,32].

In Figure 3, line-type features on TS1 and TS2 were observed. Here, HR-TEM image is given in Figure 3a, the corresponding FFT in 3b and the iFFT in 3c. On TS1 a thick twin and some narrow line-type features are observed, whereas on TS2 only narrow line-type features can be seen. In accordance, the diffraction spots for TS1 are dominating in the FFT. Furthermore, perpendicular to TS1 streaks arises. Such streaks are caused by planar defects [33]. In CoCrNi, the most plausible planar defect that can form at room temperature is a stacking fault. This leads to the conclusion that also stacking faults were formed on TS1. Also perpendicular to the TS2, streaks are observed in the FFT, which are again interpreted as stacking faults. Furthermore, these can be only observed above the twin on TS1 indicating that the stacking faults stopped at the already existing twin boundary on TS1 [22,29]. This is another indication that TS2 was activated after TS1. It can be concluded that stacking faults and twins on TS2 grow from the surface into the bulk [22]. A reason for the displacement of the twins on TS1 in different directions (Figure 1c) can currently not be offered.

To form a perfect twin lath, multiple 1/6 <112> Shockley partial dislocations need to glide on consecutive {111} slip plane. This has various geometrical consequences: One is that the ratio between the slip distance of the Shockley partial dislocation and the twin thickness normal to the {111} plane is $1/\sqrt{2} \sim 0.71$, equivalent to the shear strain of the process.

The thickness of the twin on TS2 is 39 nm, representing the sum of all {111} plane spacings as labelled in Figure 2b. The displacement of the BT was used to measure the slip distance of the Shockley partial dislocations responsible for the formation of the IT and is 37 nm. The resulting ratio is 0.95 > 0.71, which indicates that there has to be at least another mechanism releasing additional shear. The twin on TS2 shows a striped region in its upper right corner. An FFT analysis was conducted in the blue square in Figure 2b and is presented in Figure 2f. Streaks are observed and extend over the entire FFT. This means that the planar defects are most probable stacking faults and on a {111} plane [33], which is the same plane as for the twin forming Shockley partial dislocations. It is likely that the appearance of these stacking faults results in the higher shear strain caused in the twin on TS2.

In Figure 4a, a HR-TEM image of the twin intersection between the twins on TS2 and TS3 is given. The FFT is shown in Figure 4b and the iFFTs of both twins in Figure 4c+d. Figure 4 has similarities to the intersection shown in Figure 2, which are: one twin is straight (TS3), the other twin is displaced (TS2) and a striped region within the straight twin (TS3). Within the twin intersection region, line-type features are observed which seem to connect the ends of the twin on TS2 and which are marked by an arrow. An additional twin intersection mechanism, which was not discussed so far, is by the formation of secondary twins in the BT [23,34,35], whereby the BT stays straight and the IT is displaced. A schematic drawing is presented in Figure S5b. In literature, twin intersections in the same sample can be formed by different mechanisms [22,29]. A FFT of this region marked by the orange square in Figure 4a was performed to analyze if these are twins (Figure 4e). No twinning spots can be observed. Instead, streaks which are perpendicular to the line-type feature occur. This strongly suggests that the line-type features are stacking faults. As these are not dominating in the intersection region, the described mechanism by secondary twinning is unlikely. A secondary twin would result in extra spots in the FFT as the region of a secondary twin has another crystal orientation than the intersecting twins and the matrix. Based on that, the twin on TS3 should be discontinuous in the iFFT, which is not the case. We can thereby conclude, with the same arguments as for the other twin intersection, that the twin on TS2 is the BT and the one on TS3 the IT.

The striped region within the twin on TS3 was analyzed in the blue square by FFT and the corresponding image is given Figure 4f. Clear streaks are visible meaning that stacking faults were formed in almost the entire twin lath, which are on the same plane as the twinning partial dislocations discussed above. The ratio between the displacement of the twin on TS2 and the thickness of the twin on TS3 is 1.12 (calculated with the values given in Figure 4a). This is even higher than for the twin intersection in Figure 2. As the region showing the stacking faults is larger in Figure 4a than in Figure 2a, it is evident that the stacking faults carry the additional shear and were formed after the twin lath.

To summarize, twin intersections were analyzed revealing the temporal sequence of twinning. This is the first time, a deformation mechanism can be clearly traced in time under tribological load in an experiment. The first activated twin system is TS1, which is the system tilted counter clockwise towards the surface with an inclination angle of 25°. The second one is the most frequently observed twin system with an inclination angle of around 70° to the surface and labeled with TS2. The last activated system is TS3, which is the farthest from the surface. It is obvious that on the twinning systems, the resolved shear stress has to be high enough for the twins to form. This means in the given context that the temporal sequence for the crystallographic planes with the highest shear stresses has to be the same as for the twins.

This new experimental information can be used to deduce the stress field under the moving sphere in a tribological experiment.

Acknowledgments

CG acknowledges financial support by the German Research Foundation (DFG) under Project GR 4174/5-1 as well as by the European Research Council (ERC) under Grant No. 771237, TriboKey. AK acknowledges support by DFG for grant no. KA 4631/1-1. LM thanks the Deutscher Akademischer Austauschdienst (DAAD) for a PhD scholarship. Furthermore, thanks to Prof. Jens Freudenberger for experimental support and spreading joy and happiness.

Declaration of Competing Interest

The authors declare that they have no known competing financial interests or personal relationships that could have appeared to influence the work reported in this paper.

Data availability

The data that support the findings in this study are available under the link DOI 10.5445/IR/1000152864 and from the corresponding author upon request.

Figures

Figure 1. STEM images of the microstructure after the tribological experiment. a) Overview, b) and c) higher magnifications of the rectangles marked in a). TS1, TS2 and TS 3 label the three different twin systems.

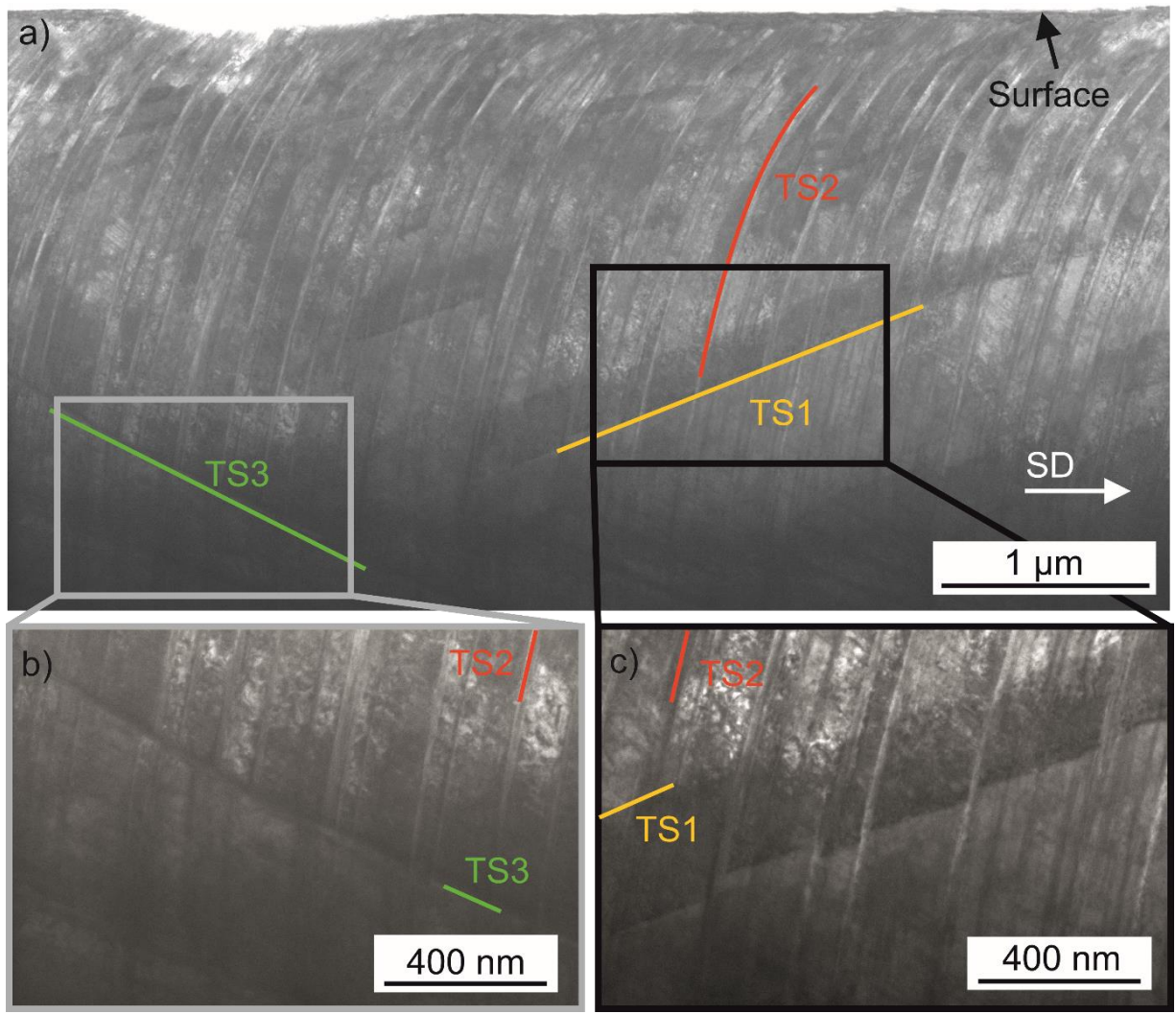

Figure 2. HR-TEM analysis of twin intersection between TS1 and TS2 from Figure 1. a) HR-TEM image, b) same as a) with additional labels, c) FFT of the whole image in a). The spots are colored in the same color as the twin they belong to. White spots belong to the matrix. d) iFFT of a) by using the twinning spots marked by the red arrows of TS2, e) iFFT of a) by using the twinning spots marked by the orange arrows of TS1 and f) FFT of the region marked by the blue rectangle in b).

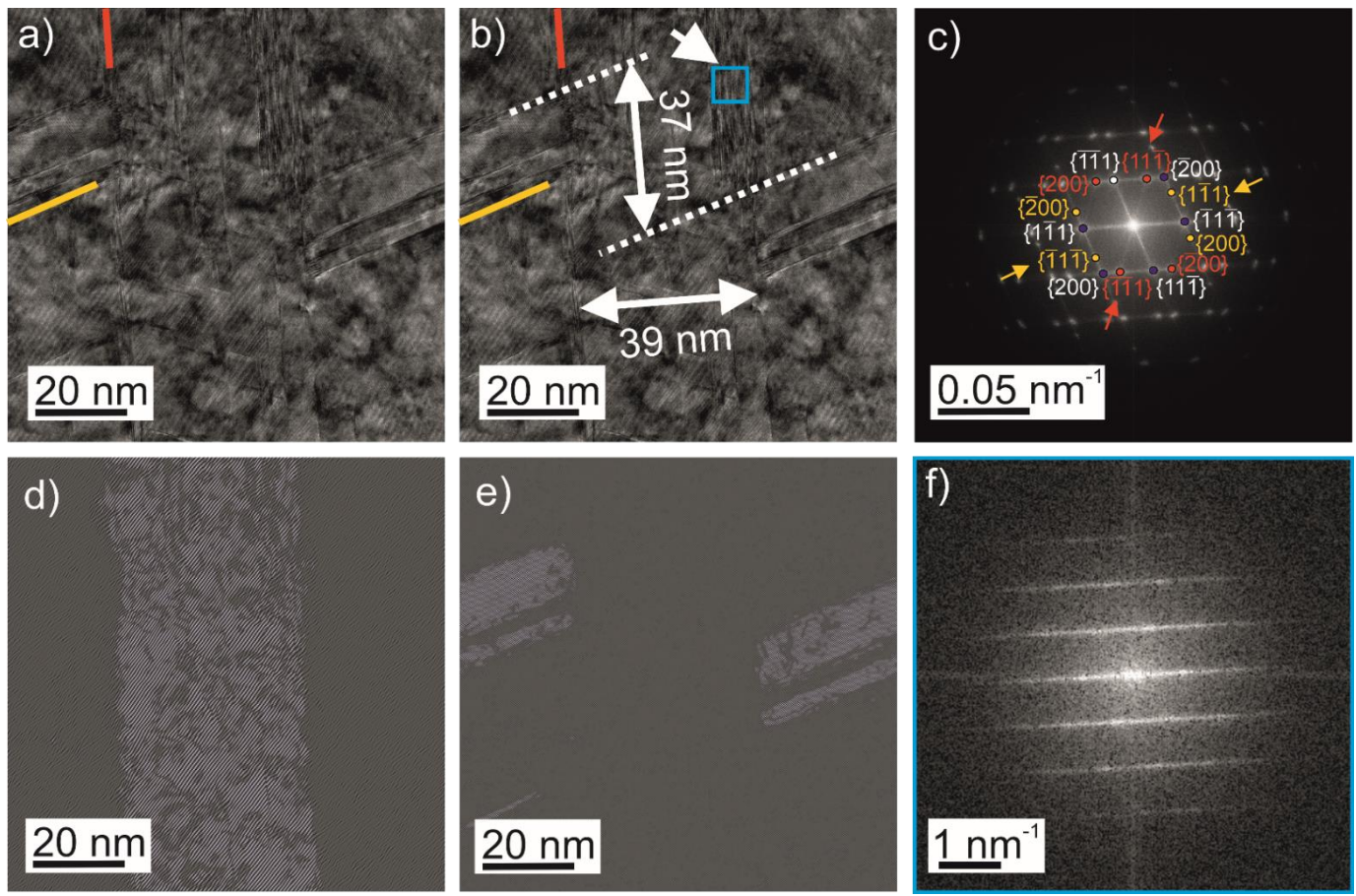

Figure 3. Interaction between TS1 and TS2. a) HR-TEM image and b) the corresponding FFT. The spots are colored in the same color as the twin they belong to. White spots belong to the matrix. c) iFFT of a) using the spots marked by the orange arrows in b).

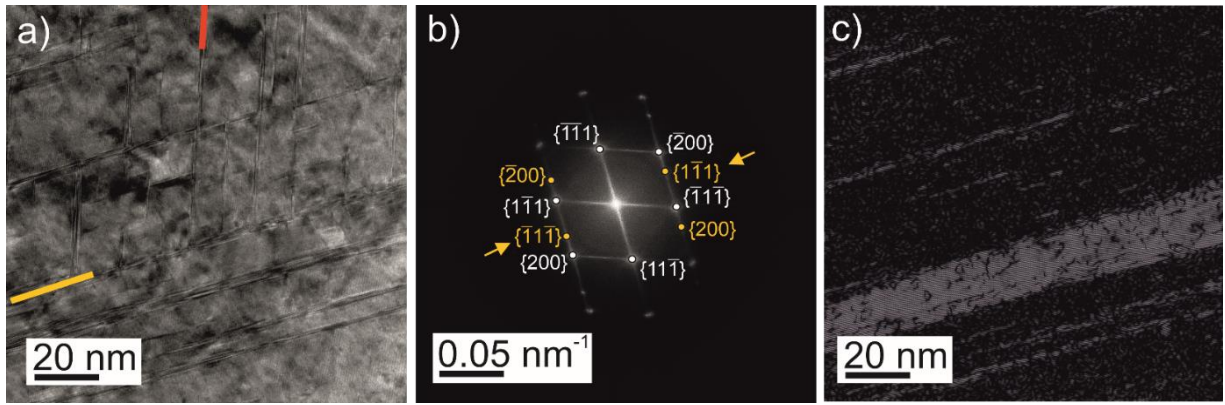

Figure 4. Twin intersection between twins on TS2 and TS3 from Figure 1. a) HR-TEM image, b) FFT of the whole image in a. The spots are colored in the same color as the twin they belong to. White spots belong to the matrix. c) iFFT of a) by using the twinning spots marked by green arrows in b) of TS3, d) iFFT of a) by using the twinning spots marked by red arrows in b) of TS2, e) FFT of the region marked by an orange-colored square in a) and f) FFT of the region marked by a blue-colored square in a).

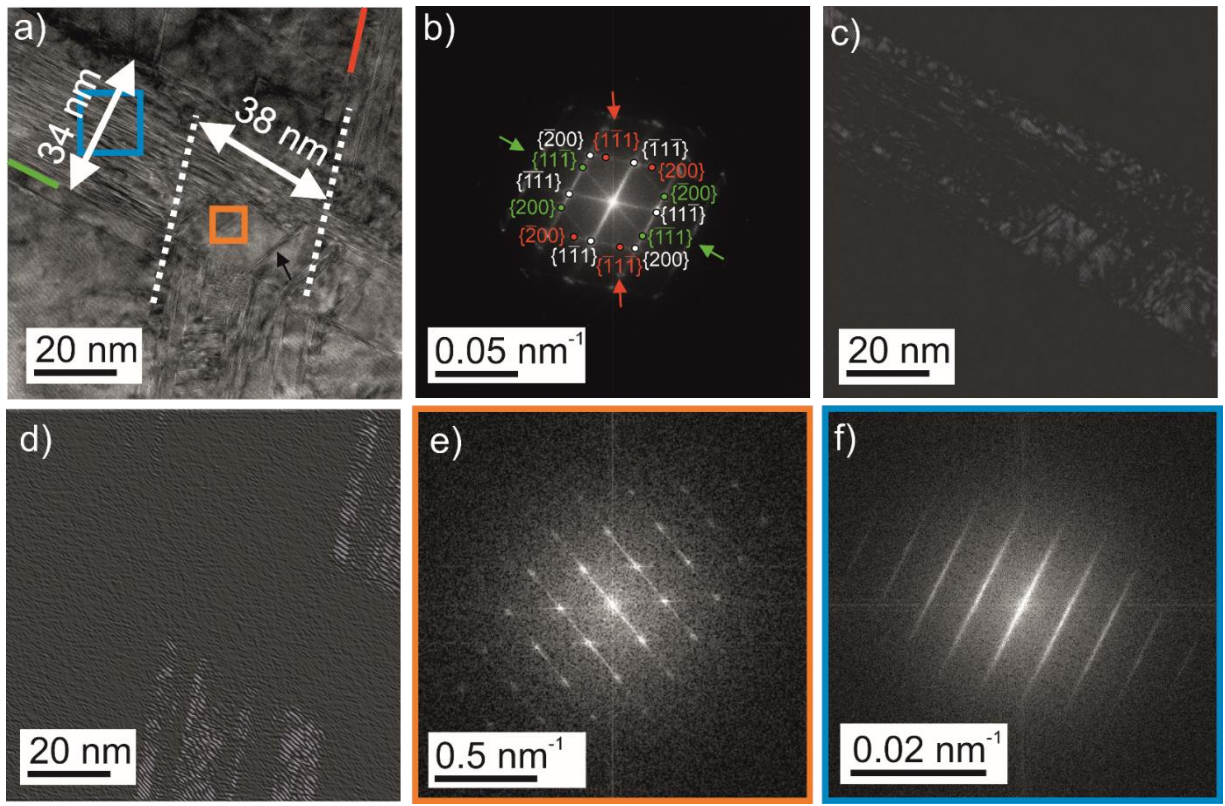

**Supplementary information**

Material manufacturing and sample preparation

Co, Cr and Ni of 99.95%, 99.99+% and 99.97% purity, respectively, were melted in an Ar atmosphere with an equimolar ratio using an AM/0.5 arc melting furnace (Edmund Bühler GmbH, Bodelshausen, Germany). The alloy was remelted for four times and finally cast into a rectangular-shaped Cu mold. Afterwards, the material was homogenized in an evacuated quartz ampule at 1200 °C for 72 h, rolled and recrystallized at 1200 °C for 1 h, resulting in a grain size of 176 ± 24 µm excluding twin boundaries measured by the linear interception method. To achieve a surface with the least defect density possible, the sample was ground down to P4000 grid, polished with diamond suspensions of 3 and 1 µm (Cloeren Technology GmbH, Wegberg, Germany) and electropolished with an electrolyte out of methanol and perchloric acid in a ratio of 9:1.

TEM investigations

Foils were cut parallel to the SD and investigated with a dual beam focused ion beam, scanning electron microscope (Helios NanoLab™ DualBeam™ 650, ThermoFisher, Hillsboro, USA) and by HRTEM (Monochromated ThermoFisher Themis-Z double corrected analytical (S)TEM). The TEM was operated at 300 kV and a Gatan Oneview IS camera was used for the image acquisition. To remove the noise in the acquired images, they were filtered using average backdrop subtraction filter (ABSF). The zone axis is in all cases [110].

Supplementary Information

Figure S1. STEM image of the as-prepared sample surface showing a defect free surface layer

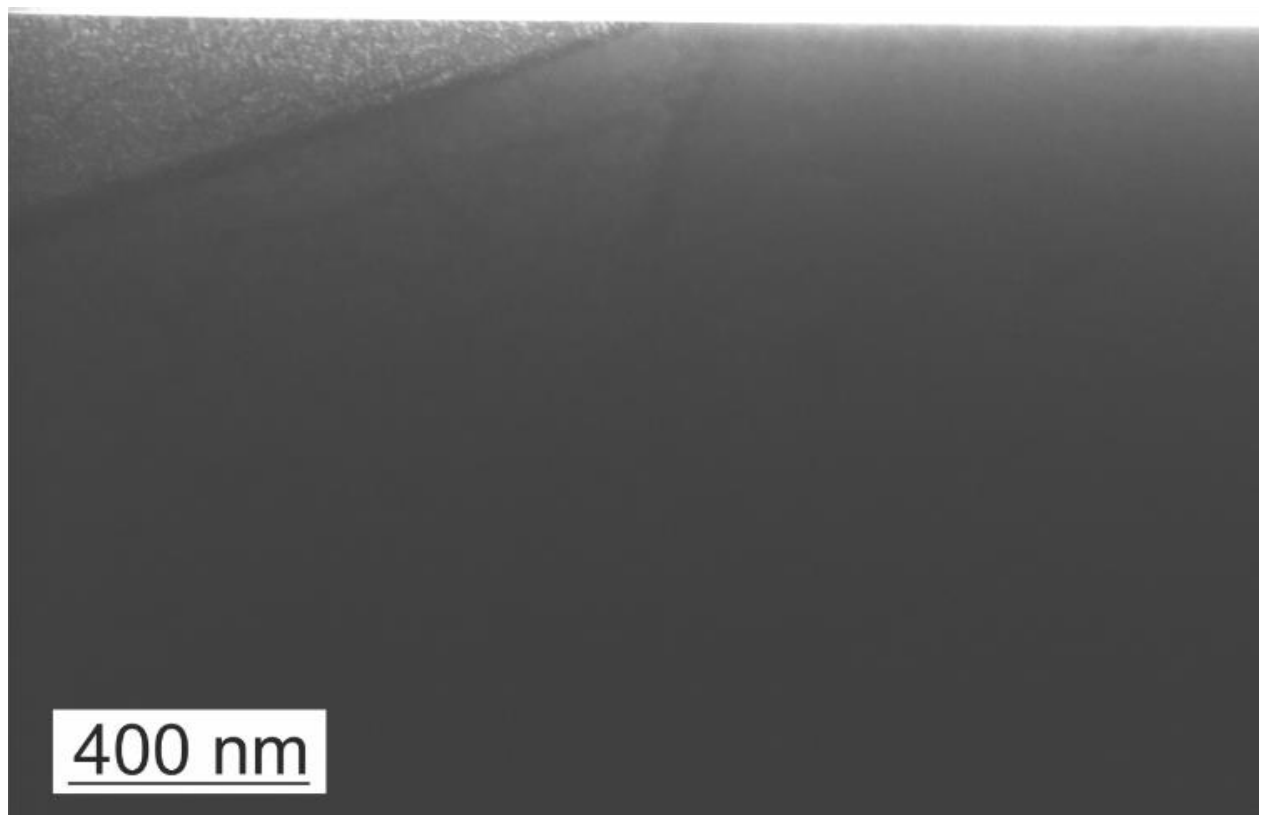

Figure S2. a) Friction coefficient over sliding distance. The grey rectangle marks the lift-out position for the TEM foils. b) SE image of the middle of the wear track and c) contact area of the SiC sphere.

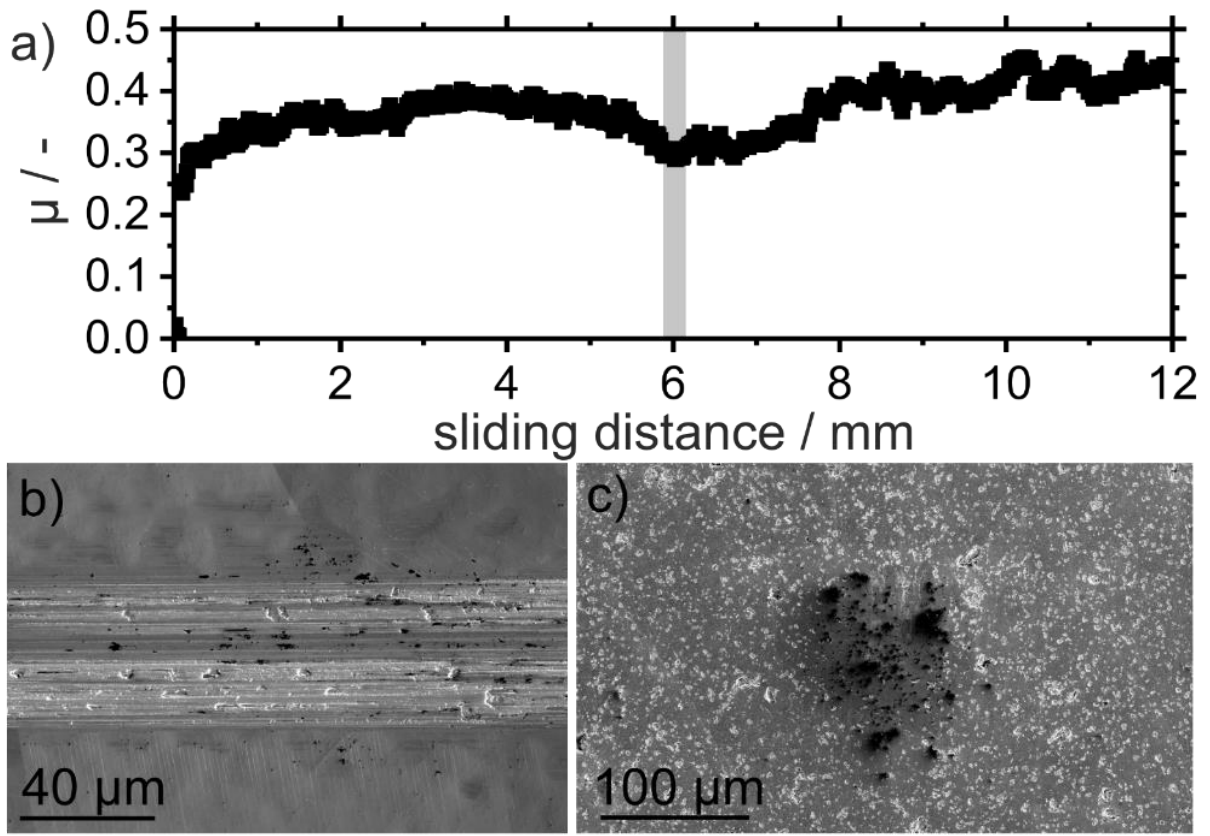

Figure S3. a) TKD measurement of the whole STEM image in Figure 1a, color-coded in normal direction. Low angle grain boundaries (<15°) are marked in green, high angle grain boundaries (>15°) in red and twins in blue ($\Sigma 3$). This measurement is stitched from three images. For this reason, there is some mismatch due to sample drift. The sliding direction is from left to right. b)-c) twinning plane analyses of the three twin systems in [111] pole figures. The red, green and orange spots correspond to the three different twin systems. The black spots mark the matrix position next to the twin. The twinning plane is the one where the matrix and the twin spot overlap.

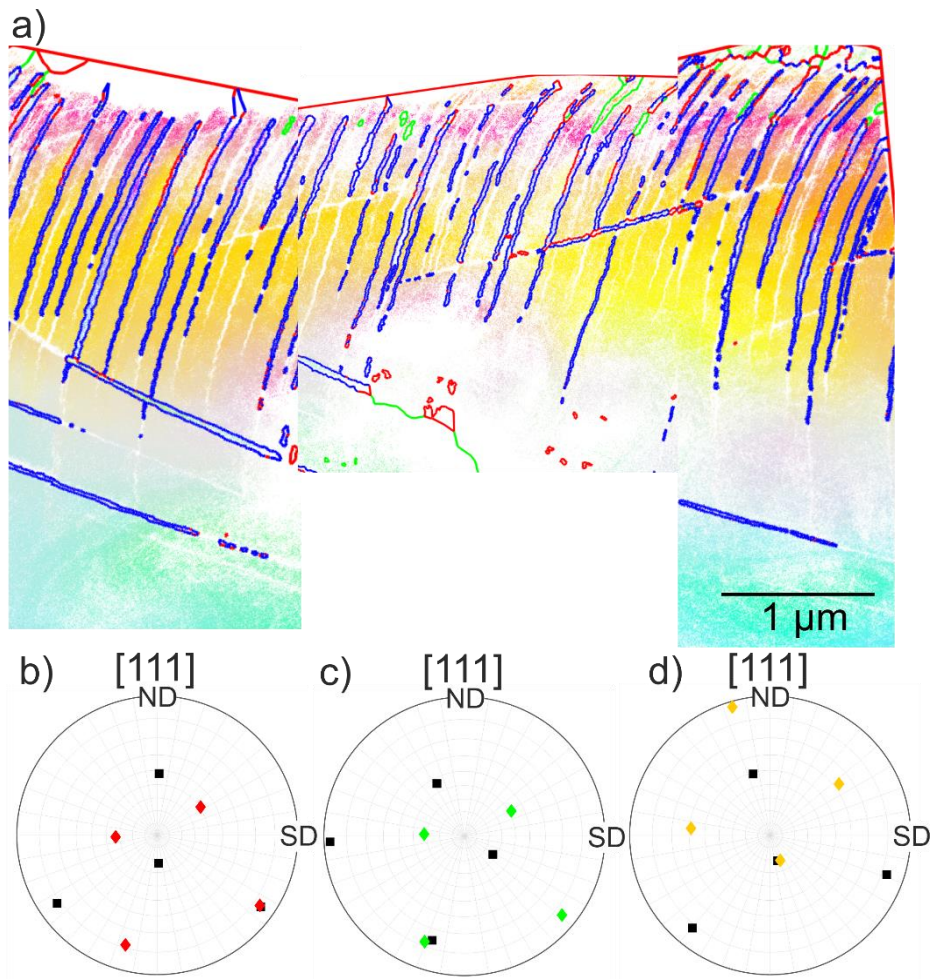

Figure S4. Twin intersection types reproduced after Figure 7 in [22]. a) Both twins have a kink and b) one twin is straight and the other is displaced.

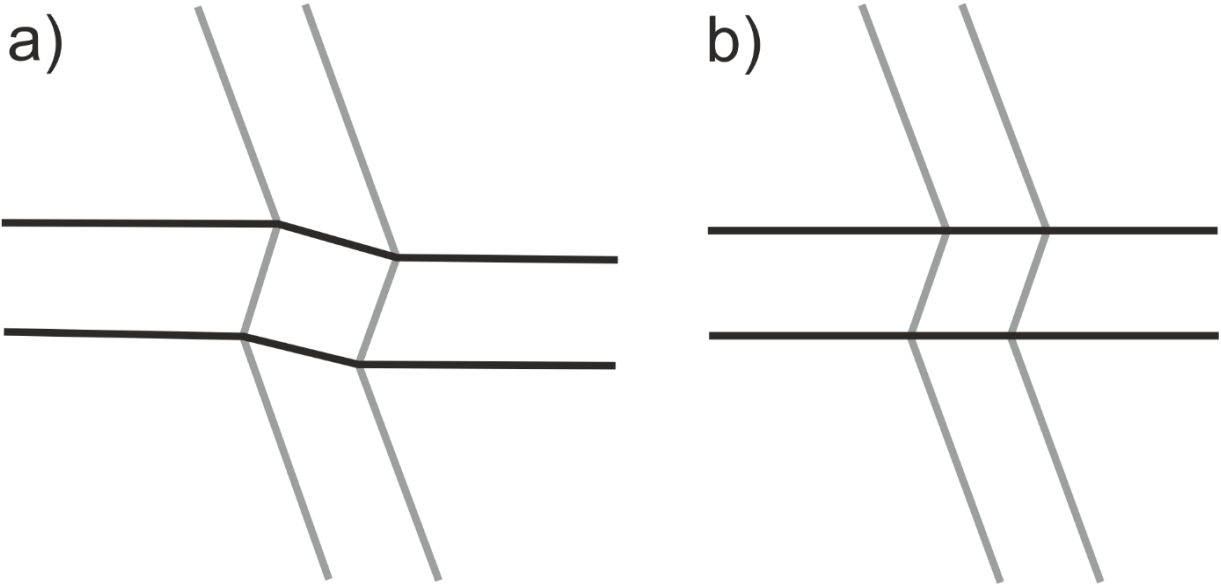

Figure S5. Atomic illustration of twin intersection models drawn with JP Minerals VESTA. a) Intersecting twins and b) intersecting twins and a secondary twin in the BT. Grey-colored atoms are matrix atoms, red-colored are BT atoms, green-colored are IT atoms and blue-colored atoms are atoms within the secondary twin within the BT.

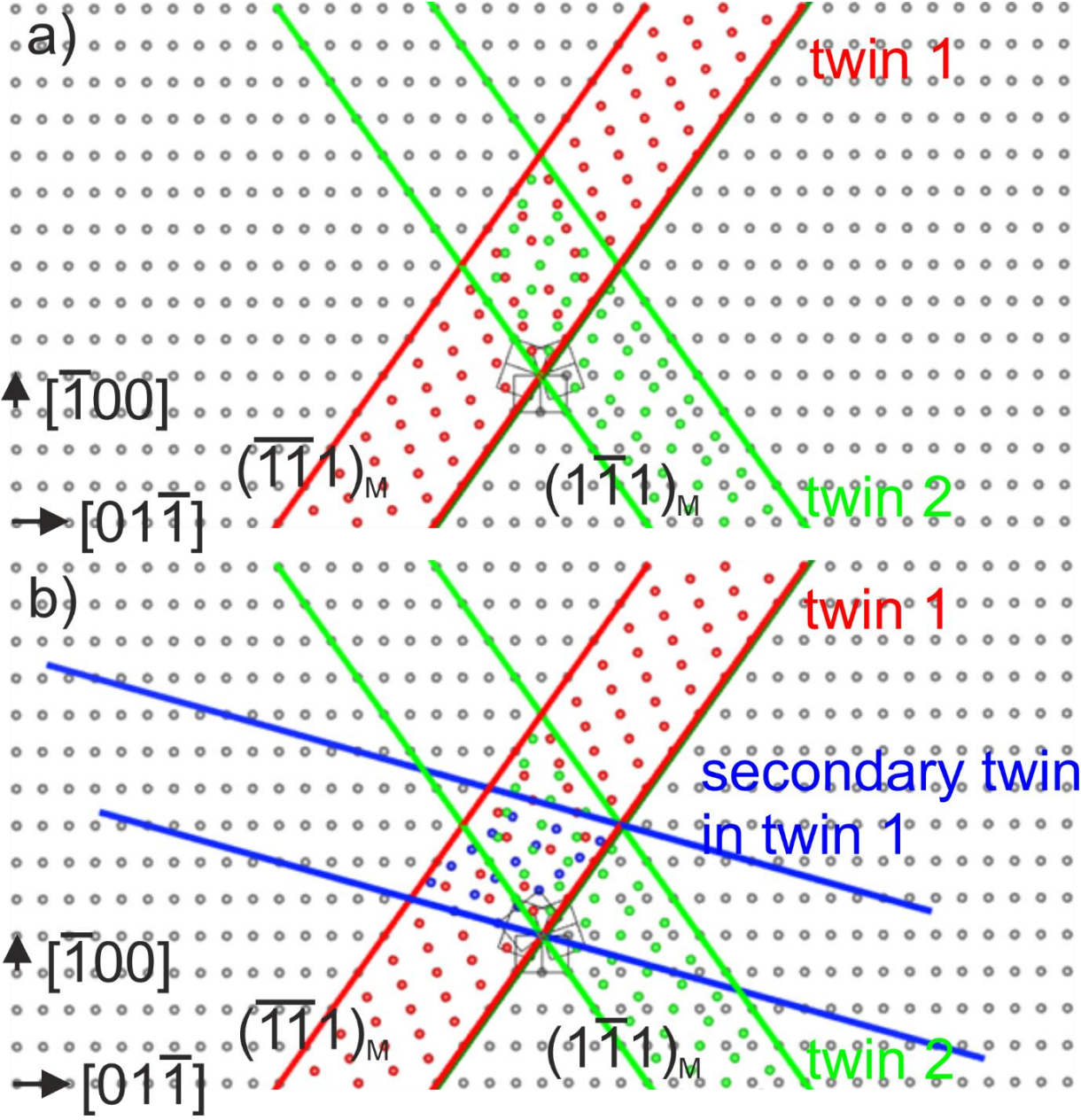